\documentstyle[12pt,aaspp4]{article}

\def\kms{\hbox {\ km\ s$^{-1}\,$}}
\def\hi {H{\small I} }
\def\hii {H{\small II} }

\lefthead{Normandeau}
\righthead{\hi towards W3}

\begin{document}

\title{Probing the Interstellar Medium using HI absorption and
emission toward the W3 HII region} 
\author{Magdalen Normandeau}
\affil{Astronomy Department, University of California,
    Berkeley, CA 94720-3411}

\begin{abstract}
\hi spectra towards the W3 \hii complex are presented and used to
probe the Galactic structure and interstellar medium conditions
between us and this region. The overall
shape of the spectra is consistent with the predictions of the Two-Arm
Spiral Shock model wherein the gas found in the --40 \kms\ to --50
\kms\ range has been accelerated by some 20 \kms\ from its rotation curve
velocity. Spin
temperatures of $\sim$100 K are derived for the Local Arm gas,
lower than found in a previous, similar study towards DR 7. For the
interarm  region, values on the order of 300 K are found, implying
a negligible filling factor for the Cold Neutral Medium ($\ll
1\%$). Some of the  
absorbing gas at velocities near --40 \kms\ is confirmed to be
associated with the \hii regions. 
\end{abstract}


\section{Introduction}

While atomic hydrogen (H{\small I}), generally seen in emission, is certainly
a pervasive component of the Galaxy, our knowledge of its optical
depth ($\tau$) and spin temperature ($T_s$) close to the Galactic plane
is sparse. The 
main reason for this is the difficulty inherent in disentangling the
emission along most sight-lines (Kulkarni \& Heiles
1988). Use of absorption towards extragalactic sources has proved
somewhat successful but has not been entirely satisfactory at low
Galactic latitudes.

The plane is peppered with bright \hii regions which could potentially
be used.
Wendker \& Wrigge (1996) studied the \hi absorption spectrum towards
DR 7 as seen using the Dominion Radio Astrophysical Observatory's (DRAO)
Synthesis Telescope (ST) and argued for the usefulness of such observations
for the determination of optical depth and spin temperature
as a function of radial velocity (or distance). 
The authors suggested that the careful study of many lines of
sight will contribute to the quasi mapping of $\tau$ and $T_s$, although 
for \hi at velocities corresponding to gas in the proximity of the \hii
regions, it is important to bear in mind that some of the absorbing
material may have been dissociated by the central stars, thus presenting
local enhancements which are not typical of the general surroundings.
This paper contributes a second line-of-sight towards a strong, extended 
continuum source within our galaxy: the W3 \hii complex, shown in
Figure~\ref{fig:W3C21}. 

The W3 region has been studied in detail repeatedly and at many
frequencies (see e.g.\ Roberts, Crutcher \& Troland 1997 for molecular line
observations; Tieftrunk et al.\ 1997 for a multiwavelength radio
continuum study; Roelfsema \& Goss 1991 for radio recombination lines;
Campbell et al.\ 1995 for a far-infrared look at subcomponents; and
Hofner \& Churchwell 1997 for X-ray). It houses active star
formation and can 
be subdivided into multiple \hii regions for which the nomenclature
varies somewhat with the type of observations. Table~\ref{tb:nomen}
defines the terminology used in this paper.
\placetable{tb:nomen}

\hi studies of sight-lines towards W3 have concentrated on searching
for atomic gas associated with the \hii regions. While the limited
spatial and spectral resolution of data presented by Sullivan \& Downes
(1973) precluded the detection of variations in \hi opacities,
observations with the Nan\c{c}ay telescope led Crovisier et al.\
(1975) to conclude that the \hi absorption near --40 \kms\ (all
velocities in this paper are with respect to the Local Standard of
Rest) is due to
atomic hydrogen related to W3. Read (1981) carried out 
a detailed study of the \hi in this region in both emission and
absorption; his data had similar spatial resolution to the DRAO data
presented here and
somewhat poorer velocity resolution (4 \kms). He suggested that
photodissociation of molecular gas near the compact \hii regions is
responsible for several observed \hi concentrations.
Goss et al.\ (1983) and van der Werf \& Goss (1990) used
the Westerbork Synthesis Radio Telescope (WSRT) and so had better spatial
resolution ($\sim30\arcsec$ and $\sim15\arcsec$) than obtainable with
the smaller DRAO array, as well as better velocity resolution (1.24
\kms\ and 1.03 \kms), but they do not provide full visibility plane
coverage and so are not sensitive to emission on all angular scales. 
They also concluded that much of the \hi at $\sim\, -40 \kms$ is due to
dissociation and delineate shells associated with various compact \hii
regions.

The above authors presented studies of the W3 complex itself and
therefore concentrated on \hi which may be related to it, i.e.\ their focus
was the \hi absorption near --40 \kms.
The present paper uses sight lines towards W3 to probe
the interstellar medium between us and this \hii region, evaluating
optical depths and spin temperatures, and placing it in a Galactic
perspective. Section 2 briefly describes the data sets which were used in
this analysis. Section 3 gives an overview of the \hi profiles towards
W3, placing their features in the context of rotation curves and
spiral shocks, as well as outlining the method used to derive the
optical depths and spin temperatures. The fourth section provides a
more detailed discussion of optical depths and spin temperatures at
different velocities. And finally, the last section summarises the findings.

\section{The data}

Radio continuum data at both 408 MHz and 1420 MHz, as well as 
21cm spectral line data
were obtained at the DRAO
as part the Canadian Galactic Plane Survey (CGPS) pilot project. The
pilot project covered an $8\arcdeg \times 6\arcdeg$ area of the sky,
encompassing 
all of the W3/W4/W5/HB3 Galactic complex. Observations 
were carried out in June, July, November and December of 1993. 
The synthesis telescope is a 7-element interferometer with four fixed
and three movable antennas, each with a diameter of $\sim$9 m. In the
course of twelve 12-hour periods, observations are taken for all
baselines from 12.858 to 604.336 m by increments of 4.286 m. At 1420
MHz, the small dish size 
results in a field of view of 78 arcmin (at 20\% attenuation of the
primary beam which is approximated by a Gaussian), the longest spacing
gives a spatial resolution of $1.00 
\times 1.00 \csc\delta$ arcmin$^2$ (EW $\times$ NS), and the shortest
baseline means that the instrument is not sensitive to structures
greater than 0.5\arcdeg\ in extent.

The spectral line data were collected
by a 128-channel spectrometer with a channel width of 2.64 \kms\ and a channel
separation of 1.649 \kms. The sensitivity at field centre for the ST data
is 3.0 K and degrades with distance from the centre as the
inverse of the primary beam. For channels where the continuum
emission from W3 is strongly absorbed there are processing artefacts
from W3.  
Information about the lowest spatial frequencies (i.e.\ large angular sizes)
was provided by the DRAO's 26m telescope, and
as a result the images are sensitive to structures
on all scales. Relevant observational parameters are outlined in
Table~\ref{tb:obs_param}. Details of the observations and data
reduction are described by Normandeau et al.\ (1997).  
\placetable{tb:obs_param}

Continuum emission was subtracted from the \hi images. While the DRAO
synthesis telescope collects continuum data at 1420 MHz using four 7.5
MHz bands, two to each side of the line frequency, the resulting image
was not used to subtract the continuum. Instead an average was taken
of channels where there was no apparent \hi, the first and last twenty
channels corresponding to velocities between +55 and +24 \kms\ and
between --137 and --154 \kms. 
This allows for a more precise subtraction of the continuum emission
for three reasons: 1) the DRAO spectrometers are equipped with
automatic level control which adjusts the gain in order to minimise
information loss due to quantization whereas the continuum system is
not, and the uncertainty of the gain correction factor would decrease the
reliability of \hi images; 2) the difference in bandwidth between a
spectrometer channel and the continuum receiver results in differing
amounts of bandwidth smearing in the images, making the subtraction of
continuum sources progressively worse with distance from field centre;
3) using the end channels insures that the visibility plane coverage
is identical for 
all the maps involved in the operation, thus allowing
a more accurate subtraction of any artefacts related to strong sources.
However, the continuum brightness temperature values used in the optical depth
calculations (see below) were from the true continuum image.

\section{Overview of \hi spectra towards W3}

The most prominent absorption features in the CGPS pilot project data 
are those associated with W3 which can be
seen out to $v_{\rm LSR} = -50 \kms$. While the profiles towards
different points within the W3 region differ significantly from each
other, mostly for velocities near --40 \kms,
they all have the same global properties which are well illustrated in
Figure \ref{fig:W3_abs_em}. This shows the average absorption profiles
associated with W3-N and W3-W for all positions where the continuum
brightness temperature is greater than 50 K. As well, an average
emission spectrum from forty nearby positions (see \S\ref{sec:calc_w3})
is plotted.

\subsection{Optical depth and spin temperature}

\subsubsection{General considerations}

Spectra towards the source (``on''), which show absorption, and towards
nearby positions (``off''), showing emission, can be combined to yield
the optical depth and spin temperature. For the on-source spectrum,
\begin{equation}
T_{\rm b,on}(v) = T_s(v)\ (1 - e^{-\tau_v}) + T_c\ e^{-\tau_v} - T_c
\end{equation}
where $T_c$ is the continuum brightness temperature of the background source.
The dependance on
frequency and therefore radial velocity, $v$, is indicated. 
The off-source
spectrum is given by the first term of the above equation:
\begin{equation}
T_{\em b,off}(v) = T_s(v)\ (1 - e^{-\tau_v}).
\end{equation}
Assuming that the off-spectrum has been carefully chosen to be representative 
of what would be seen at the position of the source if no absorption were
occurring, one can solve for the optical depth:
\begin{equation}
\tau_v = -\ln \left(1 - \frac{T_{\em b,off}(v) - T_{\em b,on}(v)}{T_c}\right).
\end{equation}
Once the optical depth has been obtained from the observed spectra and 
continuum brightness temperature, the spin temperature can be calculated
using Eq.2. It should be noted that an over- or underestimate of $T_{\em
b,off}$ will result in an over- or underestimate of $\tau$.

\subsubsection{The case of W3}

\label{sec:calc_w3}

As indicated above, the off-spectrum must be carefully chosen to
be representative of the on-source emission. This proves
difficult for sight-lines towards W3 as there is much variation.
For forty positions within a few arcminutes of W3, 20 near W3-N and 20
near W3-W as 
defined by their 50 K contours, being careful to avoid
the image artefacts and the slight absorption associated with W3-E,
for velocities ranging from +8 \kms\ to --91 \kms\ which includes all of
the Galactic emission of note, 
the standard deviation was as high as 18 K, though the average
deviation in both cases was $\sim$7 K. The average spectra, for W3-N
and W3-W, are shown in 
Figure~\ref{fig:ave_em} which also includes the
average$\pm$1$\sigma$ profiles. 

For areas where the absorption is almost total, the optical depth
calculation depends very sensitively on $T_{\em b,off} - T_{\em
b,on}$. Therefore, for the 
present purposes, whenever the value of $T_{\em b,off}(v) - T_{\em b,on}(v)$
came within $2(\sigma_{\em off}(v) + \sigma_{\em on})$ of $T_c$ it was
replaced by $T_c - 2(\sigma_{\em off}(v) + \sigma_{\em on})$ in the 
calculation, where $\sigma_{\em off}(v)$ is the standard deviation of
the appropriate off-spectrum, and $\sigma_{\em on} = 3.5 {\rm K} /
P(\theta)$, the 
uncertainty at field centre for the continuum subtracted \hi images
adjusted for the effect of the primary beam correction.
This provides a lower limit value for the optical depth. 
It affects approximately 36\% of the values of $\tau$ for W3-W from
--33.51 \kms\ to --50.00 
\kms, and $\sim$41\% of those for W3-N in this same
interval. Away from this deep trough, none of the values for W3-N
needed to be calculated in this manner,
and less than $\sim$2\% of the values for W3-W at more positive
velocities are affected.

The spin temperature calculations are highly dependant on the value of
$T_{\em b,off}$ which is the greatest source of
uncertainty. Therefore, in addition to indicating limits when only
limits were calculated for $\tau$, results for $T_s$ will be
considered trustworthy only when $T_s > 2\sigma_{T_s}$.

Plots of $\tau_v$ and $T_s(v)$ for sight lines towards W3-N and W3-W
are presented in Figures~\ref{fig:tau} and \ref{fig:Tspin}. These were
calculated using the average spectrum of all positions where the
continuum emission is greater than 50 K (i.e.\ the on-source spectra shown in
Fig.~\ref{fig:W3_abs_em}) and the average of 20 nearby spectra for the
emission (i.e.\ the emission spectra shown in
Fig.~\ref{fig:ave_em}). Calculations were also done on a pixel by
pixel basis for each region.

\subsection{Main features placed in a Galactic context}

There are three main absorption
troughs along this line-of-sight: one corresponding to Local gas, one
corresponding to the bulk of the Perseus arm and an intermediate
trough centred on $-22 \kms$.
There is \hi emission at velocities between the intermediate and Perseus
arm absorption troughs, at $v \sim -30\ \kms$. Three different
scenarios are considered to explain this and are illustrated in 
Figure~\ref{fig:diag}.

\subsubsection{Standard rotation curve without gas displacement}

\label{sec:expl1}

One possibility does not require that any of the gas be displaced
relative to the standard rotation curve, nor that the emitting gas at
--30 \kms\ be behind W3 (Fig.~\ref{fig:diag}a).

If the opacity of the gas at --30 \kms\ were low enough, there
would be sufficiently little absorption for it to be noticeable.
In this framework the values of $\tau$ calculated from the observed
spectra would imply spin temperatures in excess of 10$^4$ K for this
interarm gas, higher than expected for the Warm Neutral Medium ($T
\leq 8000$ K, Kulkarni \& Heiles 1988; direct measurements towards Cyg A
by Carilli et al.\ (1998) yielded $6000\pm1700$ K and $4800\pm1600$ K).

This feature of the spectra extends approximately from --24 \kms\ to
--33 \kms. Assuming a flat
rotation curve with an Oort constant $A = 14 \kms {\rm kpc}^{-1}$ and R$_0$ =
8.5 kpc, this implies a path length of 0.6 kpc within which there is
no cold neutral hydrogen. This seems unlikely considering that the
average off-source emission is approximately 55 K and that there is
significant absorption at more positive velocities (--20 \kms) which
also correspond to the interarm region for the standard rotation curve.

\subsubsection{Standard rotation curve with --30 \kms\ gas displaced}

In the most intuitive picture, the lack of absorption at --30 \kms\
places the related gas behind the W3 region.  

It could be argued that the --30 \kms\ gas has been
displaced in velocity (Fig.~\ref{fig:diag}b), 
that it is not following the rotation curve of
the Galaxy. In this case, the less negative velocity indicates that,
while it is approaching us, it is receding relative to the general
movement of the Galactic \hi at a velocity of at least 19 \kms, as
determined from the lower velocity edge of the --40 \kms\ absorption trough
and the upper velocity edge of the --30 \kms\ trough.

\subsubsection{TASS model, --40 \kms\ gas accelerated}

For the Perseus arm there are known to be 
large deviations from the circular motion
normally assumed for the Galaxy
(see Roberts 1972 and references therein); the location of the optical
arm and that of the radio (H{\sc I})
arm as determined using a standard Galactic velocity
curve do not coincide, with the radio arm being apparently further than the
optical arm. Roberts (1972) developed the 
``two-armed spiral shock'' model (TASS) to explain these discrepancies.
According to this model a shock develops along the inner
edge of the Perseus arm where the gas encounters a minimum in the gravitational
potential of the density wave. The shock has an amplitude of approximately
20 \kms. In this framework the main ridge of Perseus arm emission seen from
--40 \kms\ to --50 \kms, depending on the longitude
(c.f.\ Figure 6 of Normandeau et al.\ 1997), has been displaced from circular
velocity by approximately 20 \kms.
In this context the absorption gap in the W3 spectrum at --30
\kms\ would be due to the {\em undisturbed} gas behind W3, the \hii
region itself being within or just past the layer of shocked gas at
$\sim\, -40$ \kms\ (Fig.~\ref{fig:diag}c).

This is reminiscent of the streaming motion seen in other spiral
galaxies, e.g.\ M83 (Lord \& Kenney 1991) and M51 (Rand 1993 and
references therein). In M51, the strong density wave has concentrated
both the diffuse and dense gas along the inner edge of the spiral arm,
coincident with the dust lane, in the collision front. The velocity
shifts are quite large, as much as 60--90 \kms\ in the plane of the
galaxy (Tilanus \& Allen 1991). The density wave is much weaker in
M83, resulting in less pronounced streaming motions ($\sim$12 \kms\
perpendicular to the spiral arm and $\sim$0 \kms\ parallel to it; Lord
\& Kenney 1991). The dust lane 
again lies along the inner edge of the spiral arm but in this case the
molecular ridge is offset, some 300 pc downstream. Lord \& Kenney
speculate that the diffuse gas is compressed at the shock front,
producing the dust lane, but that the molecular clouds pass through
the front to form a broad distribution in the arm. 

The Perseus arm
lies somewhere between these two cases: the 20 \kms\ offset is
essentially perpendicular to the spiral arm, implying a stronger shock
than in M83 but weaker than in M51 where the perpendicular component
is  approximately 64 \kms (Lord \& Kenney 1991).
Heyer \& Terebey (1998) studied the CO and infrared emission in the
W3/W4/W5 region and found the bulk of the CO emission to be at
velocities near --45 \kms. They estimate that the minimum transit
time for the interarm region requires an exceedingly long cloud
lifetime (3--6 $\times$ 10$^9$ yr), and, when combined with the
arm-interarm contrast which they measure to be 28:1 for CO, this implies that
the gas which enters the spiral arms is in the atomic phase. All this
is more akin to the situation in M51 than in M83, 
with a pile-up of molecular material along the shock front,
though the shock is weaker for the Perseus arm than in M51. It should
be pointed 
out, however, that Heyer \& Terebey discounted the possibility that
the gas at --40 \kms\ was in fact showing streaming motion.

Frail \& Hjellming (1991) presented an absorption
spectrum towards LSI+61$\arcdeg$303, a Perseus arm object located just
east of W4. It 
is very similar in appearance to the one presented here for W3,
and they also called upon the TASS model to explain their observations.
A large scale phenomenon such as a spiral density wave seems the most likely
explanation for the similar velocity displacement of gas separated by
some 78 pc (assuming a distance of 2.3 kpc). 
Frail \& Hjellming also show a spectrum toward an extragalactic
source, BG 0237+61 
which is 15 arcmin from LSI+61$\arcdeg$303, where there is
absorption in the --20 \kms\ to --40 \kms\ velocity interval; this
argues against the possibility outlined in \S\ref{sec:expl1}.
For these reasons, the 
explanation involving the TASS model is the one favoured here.

\section{Discussion of optical depth and spin temperature}

\subsection{Absorption near 0 \kms}

Of the \hi studies towards W3 mentioned in the introduction, only the
earlier ones cover a wide-enough velocity range to include the absorption
by Local gas. Crovisier et al.\ found that the optical depths and
widths of the troughs for W3 N and W3``main'' were comparable. The
present data show slight differences. The optical depth towards W3-W
rises and falls 
smoothly, attaining at maximum of $0.66\pm0.08$ at $-0.53 \kms$,
whereas towards W3-N there is a brief plateau, extending over some 8
\kms, at a level of 0.6--0.7. Read also found a peak optical depth of
$\sim$0.6. Within the two subregion there is little variation.

The spin temperature towards W3-N
attains slightly lower values than towards W3-W and remains at this
level over a wider velocity interval, corresponding to the 8 \kms\
plateau mentioned above. Both
sight-lines show slightly lower $T_s$ (as low as $99\pm19$ K towards W3-N and
$115\pm20$ K 
towards W3-W) than found by Wendker \& Wrigge towards DR 7 (generally
around 140 K). Considering the
uncertainties, the discrepancy is not large. Nonetheless
a possible explanation for this could be that the presence of 
relatively small, colder \hi ``clumps'' would have a much
greater impact on the spin temperatures measured towards W3 than
towards DR 7. This is because for each channel only a mean brightness
temperature is measured 
and, due to Galactic rotation,
the same channel width (velocity width) corresponds to a greater
path length for the DR 7 case than for W3. For DR7 sight-lines
each channel would therefore
include contributions from much
warm gas as well as from the postulated small, cold clumps.
As an illustration, for 
$v_{\rm LSR} = -10 \kms$ and assuming a flat
rotation curve with an Oort constant $A = 14 \kms {\rm kpc}^{-1}$ and R$_0$ =
8.5 kpc, one finds d$_{\rm kin}$ = 4.5 kpc for the longitude of DR 7,
but only 0.7 kpc for sight-lines towards W3\footnote{Kinematic
distances can be calculated for this local gas because it is
unaffected by the Perseus streaming motions.}.

\subsection{Between the 0 \kms\ and --20 \kms\ absorption troughs}

In the interval between the troughs at 0 and --20 \kms,
Figure~\ref{fig:W3_abs_em} indicates that there is some absorption,
though not to the extent seen in the deep troughs. The optical depth
calculations reach minima of $0.14\pm0.03$ towards W3-W and
$0.08\pm0.04$ towards 
W3-N. It is reasonable to wonder if this is truly due to interarm
absorption or if it is simply attributable to the overlap of wings from the
distributions corresponding to the 0 and --20 \kms\ troughs. Fitting a
Gaussian to each of these shows that their wings cannot account for the
amount of absorption seen at intervening velocities.

For both sight-lines, the spin temperature increases upon leaving the
Local Arm. The rise and fall of $T_s$ in this interarm region is
fairly smooth towards W3-W, reaching $286\pm79$ K when using the average
absorption spectrum. 
Towards W3-N the interarm spin temperature shows
a double peak, though it is smooth within uncertainties. 

The above temperatures are, of course, weighted mean values for a given
channel.
They correspond to what Kulkarni \& Heiles (1988) dubbed the {\em
naively derived spin temperature}. For a single,
isothermal cloud, it is equal to that cloud's spin
temperature, however for the more complicated and realistic case where
there are contributions from two optically thin components, it becomes
the column-density-weighted harmonic mean temperature. Assuming the
canonical values of 80 K and 8\,000 K for the Cold Neutral Medium
(CNM) and the Warm Neutral Medium (WNM)
respectively (Kulkarni \& Heiles 1988) and using 300 K as the value
for the weighted harmonic mean spin temperature, one derives a column
density that is three times as high for the WNM as for the CNM. From
the measured values of the spin temperature and the optical depth in
the --8.78 \kms\ to --13.72 \kms\ velocity interval, the total column
density of the \hi $3.4 \times 10^{20}$ cm$^{-2}$, implying a column
density of $8.4 \times 10^{19}$ cm$^{-2}$ for the CNM and $2.5 \times
10^{20}$ cm$^{-2}$ for the WNM with the above assumptions. If one then
uses as a path length the difference between the kinematic distances
associated with the velocities given above (0.63 kpc and 0.96 kpc), one
finds mean densities of 0.08 cm$^{-3}$ for the CNM and 0.24
cm$^{-3}$ for the WNM. The ISM pressure in the plane is thought to be
approximately $\frac{P}{k} \sim 4000$ cm$^{-3}$ K (Kulkarni \& Heiles
1988), implying $n_{\em WNM} = 0.5$ cm$^{-3}$ and $n_{\em CNM} = 50$
cm$^{-3}$. Taking the ratio of the mean densities for the interarm
velocities considered here and the expected densities in the plane,
one finds volume filling factors $\sim$50\% for the WNM and $\sim0.2\%$
for the CNM. The WNM filling factor is comparable to the value of
$\sim$40\% which Carilli et al.\ (1998)  found for the average of two
interarm regions along a line of sight towards Cyg A, though using only their
data for the interarm region between the local gas and the Perseus arm
results in a somewhat lower value, closer to 30\%.

The CNM and WNM filling factors calculated above are very uncertain,
relying as they do on many assumptions.  
The temperatures adopted for the two components may be incorrect. 
Carilli et al.\ measured
temperatures for the WNM down to 4800$\pm$1600 K. While using this
lower value would not noticeably change the mean densities derived for
the two components in the interarm region, it would greatly
affect the expected density for the plane and would 
imply a filling factor of $\sim$30\% for the WNM. The CNM filling
factor would, of course, be unaffected.
The temperature chosen for the CNM has a
greater impact on the implied relative column densities of the two
phases; for $T_{\em CNM} = 20$ K, the filling factor of the WNM
becomes $\sim$60\%, while for the CNM it is then below 0.1\%. 
Deriving path lengths from the kinematic distances can only give rough
estimates. Small differences in this number will result in widely
differing estimates of the filling factor for the WNM ($\sim$30\% for
$\ell = 0.5$ kpc and $\sim$80\% for $\ell = 0.2$ kpc), though the
CNM filling factor remains at the 0.1--0.2\% level.
Clearly, the greatest uncertainty comes from the densities assumed for
the two components in the plane. These are affected by the
temperatures as noted above, but also depend on the assumption
that the phases of the ISM are in pressure equilibrium\footnote{The
expression for the pressure was derived for the Warm Ionized Medium
and can only be applied to other components if there is pressure
equilibrium}, as well as on the reliability of the equation.
The quantitative results quoted in the preceeding paragraph are
therefore not to be blindly trusted, particularly for the WNM, however
the calculations clearly and reliably show that the CNM occupies a
negligible fraction of the interarm region. Nonetheless this small
amount of cold gas dominates the absorption signal. 

\subsection{Absorption near --20 \kms}

Crovisier et al.\ interpreted their observations at these velocities
as due to the presence of a large inhomogeneous cloud at a distance of
approximately 1.5 kpc. In the context of the TASS model favoured here,
the difference between sight-lines towards W3-N and W3-W is due to
inhomogeneities in the gas on the near side of the Perseus Arm.

Towards W3-N the average optical depth rises and
falls quickly, reaching only $0.49\pm0.07$ at --20.32 \kms. For the W3-W
line-of-sight, the rise is more gradual than the decline and the
maximum attained is much greater, $1.66\pm0.26$ at --20.32 \kms. Read found
a peak optical depth of $\sim$1.4 for this trough, in agreement
with the value above.
Accordingly, sight-lines towards W3-W reach lower spin temperatures
($69\pm13$ K for the average profile) than do those towards W3-N
($119\pm23$ K for the average profile).

The optical depth is fairly uniform over the individual regions. A few
anomalously high values (up to 2.6) do result from the calculations
but these are correlated with steep gradients in the continuum image and are
therefore likely to be artefacts. Two factors which could have affected
these pixels are: 1) slight misalignments of the continuum and spectral
line images, which would have a great impact on calculations in these
steep gradient regions, are not impossible as the continuum image was
selfcalibrated; 2) the gridding of the continuum image and of the spectral
cube were done separately and both have been regridded, which procedure
could have caused mismatches between the two data sets in steep gradient
regions. Because the affected region is very near the field centre,
bandwidth smearing is not a likely cause of these errors. 

\subsection{Absorption near --40 \kms}

As indicated in the introduction, the absorption at velocities near
--40 \kms\ has been studied with the WSRT at higher spatial and
spectral resolutions (Goss et al.\ 1983, and van der Werf \& Goss 1990). A
new detailed analysis of the \hi associated with the various compact
\hii regions using the DRAO data is therefore not warranted,
but a few aspects shall be addressed and
discussed. 

For the W3-N region, the lower limit for the maximum value using the
average spectrum is 1.6. 
The optical depth varies little over the region; for the pixel by pixel
calculations, 
a lower limit of 2.9 on the maximum optical depth was found at
a velocity of --40.11 \kms.

For this region Goss et al.\ calculated values up to 2.5
whereas Read found $\tau$ up to $\sim$8. The former are inconsistent
with the present data. The latter may be overestimated due to poor
continuum subtraction leading 
to an underestimate of $T_{\em b,on}$ (Read did not have channels
perfectly devoid of continuum at his disposal for continuum
subtraction and estimated that this may have resulted in a 10 K
oversubtraction). 
It is possible that the Westerbork results have underestimated the
optical depth. The WSRT has a shortest baseline of 36 m and, as a result,
the largest scale structure that can be imaged is $\sim$10 arcmin in
extent. This filtering property of the array highlights
small scale knots, and therefore would not affect the absorption
spectra because the structure of the absorbing sources is on scales
smaller than 10 arcmin (c.f.\ Figure~\ref{fig:W3C21}). If all the \hi
emission was on scales greater than 10 arcmin then there would be no
surrounding \hi emission which would need to be accounted for through
an off-source spectrum. However if the interferometer has not filtered out
all of the \hi emission, then neglecting it, as did
Goss et al., will result in an underestimate of $\tau$.
The W3-N opacity images published by van der Werf \& Goss show values of
$\geq 4$; their calculations were done differently than those of Goss
et al., circumventing the possible difficulty with $T_{\em b,off}$
(see van der Werf \& Goss 1989 for details of the method).

It should also be noted that for these velocities, it is likely that
the optical depths derived here are also underestimated. This is
because there is probably local \hi associated with the W3 region
itself which will not be accounted for in the ``off'' spectrum. The
latter will therefore contain less \hi emission at these velocities
than would a spectrum towards the source if there were no absorption.

Spatial variations are much more marked in W3-W which, contrary to
W3-N, is made up of several distinct \hii regions. In particular, the
highest values are attained towards W3 core (lower limit of 4.7 at
--38.46 \kms\ whereas the lower limit for the maximum from the average
spectrum is 2.4), and
lesser but marked enhancements are seen towards W3 K and W3 J as well
as in the southern section of NGC 896. Again Goss et al.\ obtained
slightly lower
values, with a maximum of 4.0 being detected towards W3-A and W3-B.
However van der Werf \& Goss found $\tau > 5.0$ towards all continuum
sources and explained the discrepancy in terms of beam dilution;
this reconciles the two Westerbork data sets, though beam 
dilution should cause the optical depth evaluated with the DRAO data to
be even lower. 
As for Read, once 
again he quotes a higher value (lower limit of 9.0), but the same
warning applies as for W3-N. 

The average spin temperature is fairly constant over the entire
velocity range of the absorption trough, for both W3-N and W3-W,
at somewhat less than 100 K. This temperature is consistent with
expected values for cores of Giant Molecular Clouds (20 -- 100 K, Turner 1988).

\section{Summary and conclusions}

\hi spectra towards W3-N and W3-W for velocities ranging from +55
\kms\ to --154 \kms\ have been presented. These were combined with an
average spectrum for nearby positions to calculate the optical depth
and spin temperature for velocities where there was absorption of the
continuum emission. 

There is a lack of absorption around --30 \kms\ which may be indicative
of temperatures in excess of $10^4$ K in the interarm region or which
might correspond to gas behind the W3 region even though there is
absorption out to --50 \kms. In the latter case, it could be
that the --30 \kms\ gas has been displaced by $\geq 19 \kms$ from the
standard rotation curve, or that it is the gas showing absorption near
--40 \kms\ that has been accelerated by a spiral shock in accordance
with the Two Armed Spiral Shock model (Roberts 1972). Considering that Frail \&
Hjellming (1991) observe a very similar spectrum towards
LSI+61$\arcdeg$303, a source east of W4, the 
explanation wherein the --30 \kms\ gas has been displaced 
seems the least probable. The explanation involving the TASS model 
is favoured here
because of the unlikely absence of cold \hi over the velocity span of
the --30 \kms\ feature.

For the Local Arm, the sight-lines presented here yield lower spin
temperature values than reported by Wendker \& Wrigge (1996) towards DR
7. This discrepancy may be due to the longer path length
corresponding to each channel width for the earlier study. Additional
investigations of this nature, towards other Galactic plane \hii regions seen
in the CGPS data, will clarify the matter. For the interarm region,
values on the order of 300 K are found, from which one can estimate
volume filling factors of $\sim$50\% for the WNM and $\ll 1\%$ for the
CNM; the calculations require many assumptions and the number
quoted for the warm \hi cannot be said to be reliable, but the result
for the CNM filling factor is quite robust. The --20 \kms\
absorption 
trough which shows lower temperatures is part of the Perseus arm in
the TASS model, not the interarm region. 

The study of this second line of sight towards a Galactic plane \hii
region, following on work by Wendker \& Wrigge towards DR7, confirms
the usefulness of such studies both in determining characteristics of
the ISM and in examining elements of Galactic structure. 
The many \hii regions within the 73\arcdeg\ longitude
span of the CGPS should help us map the temperature and optical
depth in the Galaxy.

\acknowledgements

The author is grateful to H.J.\ Wendker and C.\ Heiles 
for useful comments on previous drafts, as well as to an anonymous referee for
comments helpful in the preparation of the final manuscript.  
The Dominion Radio Astrophysical
Observatory's synthesis telescope is operated by the National Research
Council of Canada as a national facility. 
The Canadian Galactic
Plane Survey is a Canadian project with international partners, and is
supported by a grant from the Natural Sciences and Engineering Research
Council of Canada.  

\clearpage

\begin{deluxetable}{l c c l}
\scriptsize
\tablewidth{0pt}
\tablecaption{Components of the W3 region \label{tb:nomen}}
\tablehead{
\colhead{Name} & \colhead{$(l,b)$} & \colhead{Approx.\ size} &
\colhead{Description}
}
\startdata
W3 North (W3-N) & (133.78, +1.42) & $9.5' \times 8.5'$ & Evolved HII region \nl
W3``main'' & (133.79, +1.18) & $16.5' \times 14.0'$ & All the
	bright emission south of W3-N \nl
W3 East (W3-E) & (133.81, +1.18) & $9.5' \times 14.0'$ & Part of W3``main''.
	Fan-shaped, lower \\ & & & intensity evolved HII region \nl
W3 West (W3-W) & (133.72, +1.17) & $6.0' \times 10.0'$ & Part of
	W3``main''. Bright western emission \nl
W3 core & (133.71, +1.22) & $4.5' \times 3.5'$ & Part of W3-W. Bright
	northern sources \nl
W3 A+B & (133.72, +1.22) & $< 1'$ & Part of W3 core. Unresolved
	HII regions \nl
W3 H+C+D & (133.69, +1.22) & $< 1'$ & Part of W3 core. Unresolved
	HII regions \nl
W3 K & (133.73, +1.18) & $< 1'$ & Part of W3-W. Compact HII region
	\\ & & & south of W3 core \nl
W3 J & (133.70, +1.17) & $< 1'$ & Part of W3-W. Compact HII region
	\\ & & & south of W3 core \nl
NGC 896 & (133.70, +1.14) & $4.5' \times 6.0'$ & Part of W3-W. Large ring
	south of W3 core \nl
\enddata
\tablenotetext{a}{Sizes are approximate, being dependant on the threshold used
        to determine the boundary of a particular region. They are
        given here to help the reader associate the names listed with
        the features seen in Fig.~\ref{fig:W3C21}, and correspond
        roughly to the lowest closed contour $\geq 20 {\rm K}$ around a given
        source. It should be noted in particular that W3-E could be
        said to extend further east than suggested by the dimensions
        given above.}
\tablenotetext{b}{In this paper, W3 East (or W3-E) is used to refer to the 
	fainter emission east of W3``main''; this is the standard naming
	convention for low and intermediate resolution centimeter observations.
	It is important to point out that in infrared studies, W3 East often 
	refers to the eastern of the two far-infrared sources in W3``main''.
	Also, higher resolution radio studies sometimes refer to a component
	of W3``main'' as W3 E.}
\end{deluxetable}

\clearpage

\begin{deluxetable}{l l l}
\tablewidth{0pt}
\tablecaption{Observational parameters for the spectral line data
\label{tb:obs_param}}  
\tablehead{
\colhead{Parameter} & \colhead{Value} 
}
\startdata
ST primary beam & Gaussian, FWHM = 103 arcmin \nl
ST baselines & 12.858 m to 604.336 m, increment = 4.286 m \nl
Spatial resolution & $1.00' \times 1.10'$ (EW $\times$ NS) \nl
Polarisation & RR \nl
Bandwidth & 1 MHz \nl
Central velocity (LSR) & --50.0 \kms \nl
Velocity coverage & 211 \kms \nl
Channel width & 2.64 \kms \nl
Channel separation & 1.65 \kms \nl
ST sensitivity at field centre & 3.0 K \nl
\enddata
\end{deluxetable}

\clearpage

\plotone{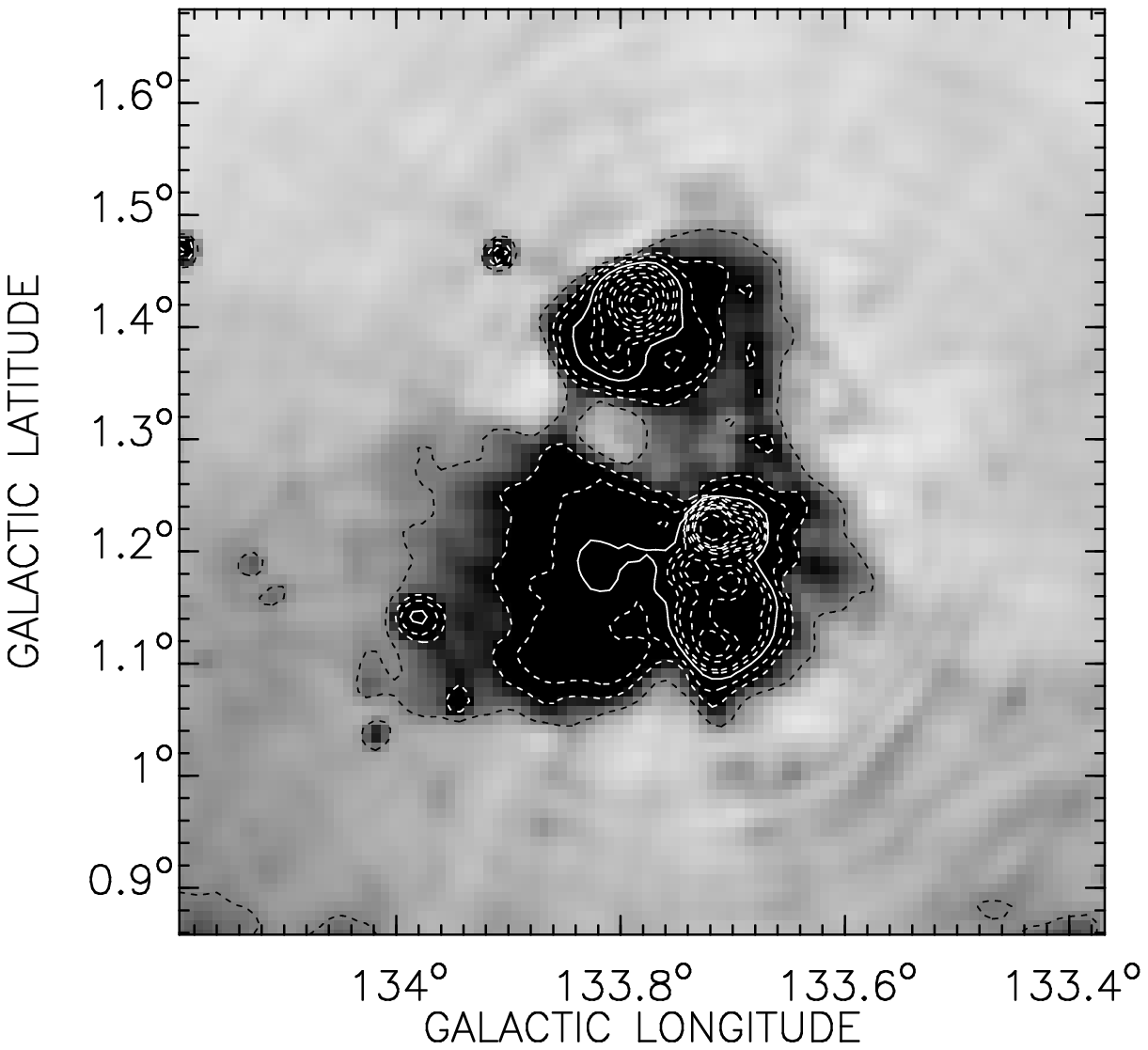}

\figcaption[Normandeau.fig1.ps]{The W3 region at 1420 MHz. 
	The greyscale varies from 0 K to 20 K. The solid line contour
        is at 50 K and the dotted contours are at 10, 20, 30, 70, 90,
        150, 200, 300, 400, 500 and 1000 K.\label{fig:W3C21}}

\clearpage

\plotone{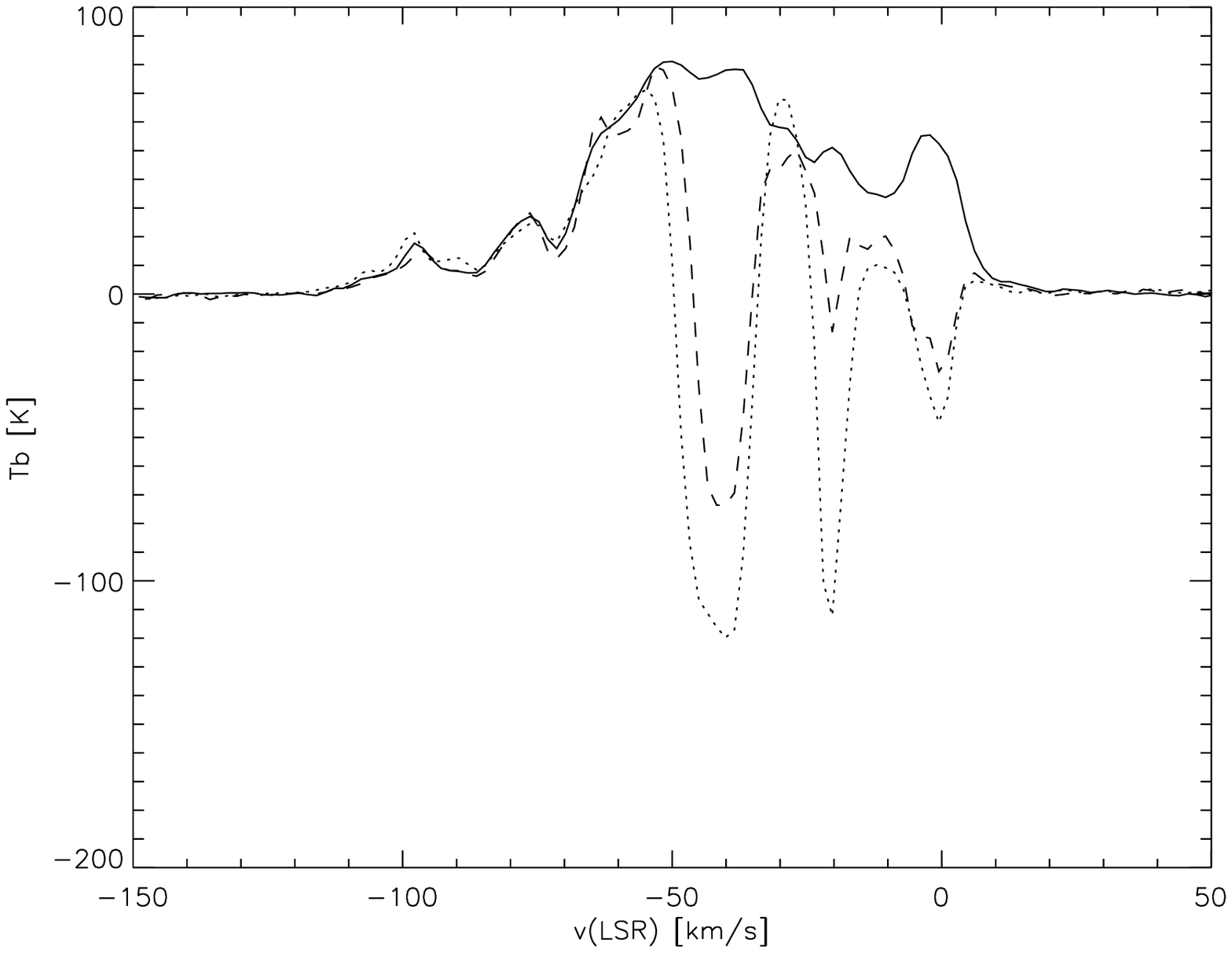}

\figcaption[Normandeau.fig2.ps]{Absorption of W3-N and W3-W.
	\hi spectra along lines of sight towards W3-N (dashed line)
        and W3-W (dotted line) are plotted along with the average emission
        spectrum of the surrounding gas (from an average of 40 nearby
        positions, see text).
	The absorption spectra were obtained by averaging the \hi
	images for all pixels for which the continuum values were
	greater than 50 K, at the positions of W3-N and
	W3-W.\label{fig:W3_abs_em}} 

\clearpage

\plotone{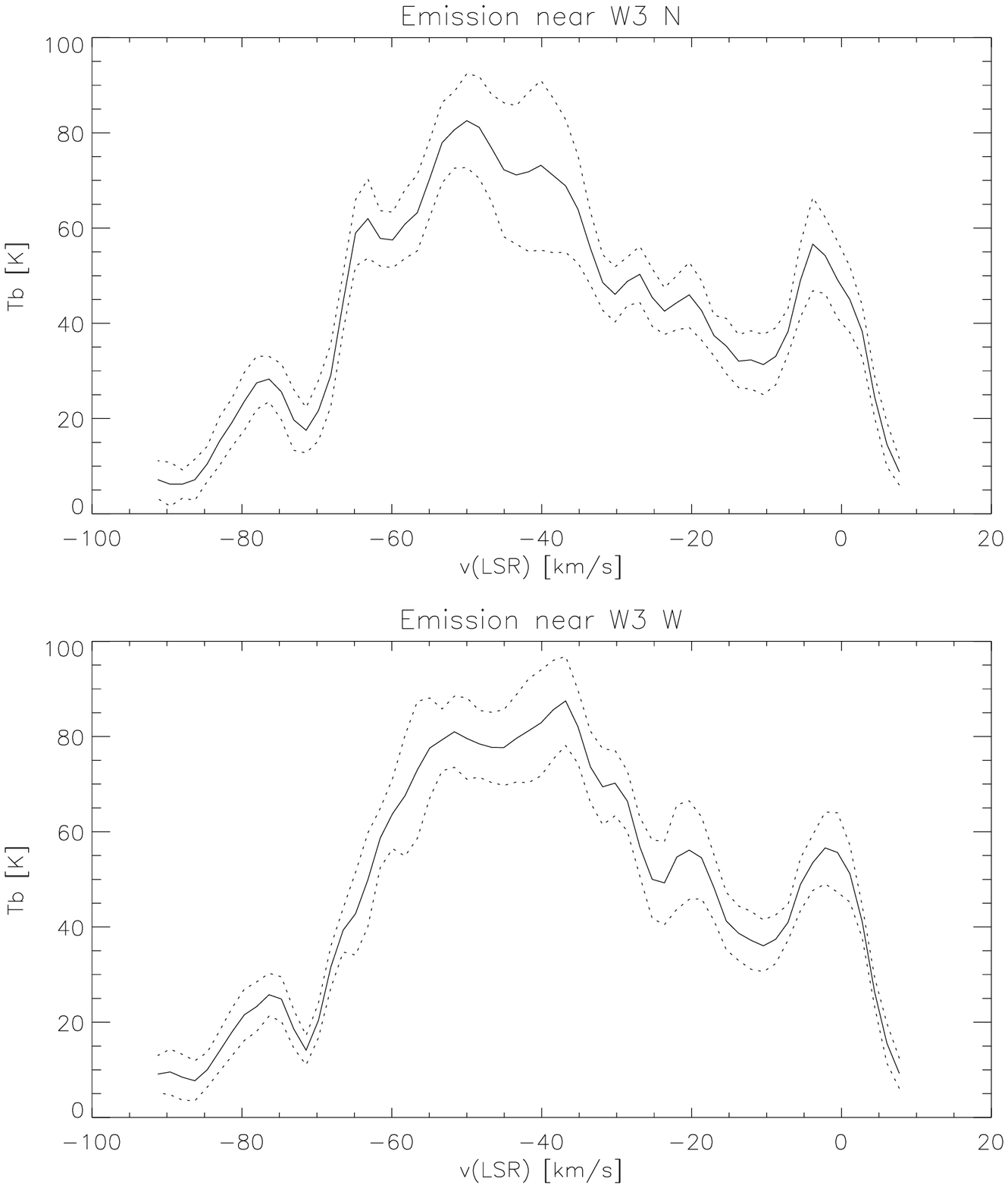}

\figcaption[Normandeau.fig3.ps]{Average emission spectra towards W3-N and W3-W.
	A sample of twenty nearby positions for both W3-N and W3-W
	were used to derive the average emission spectra shown by the
	solid lines. The dotted
	lines are at plus and minus $1\sigma$ from these average
	curves.\label{fig:ave_em}}

\clearpage

\plotone{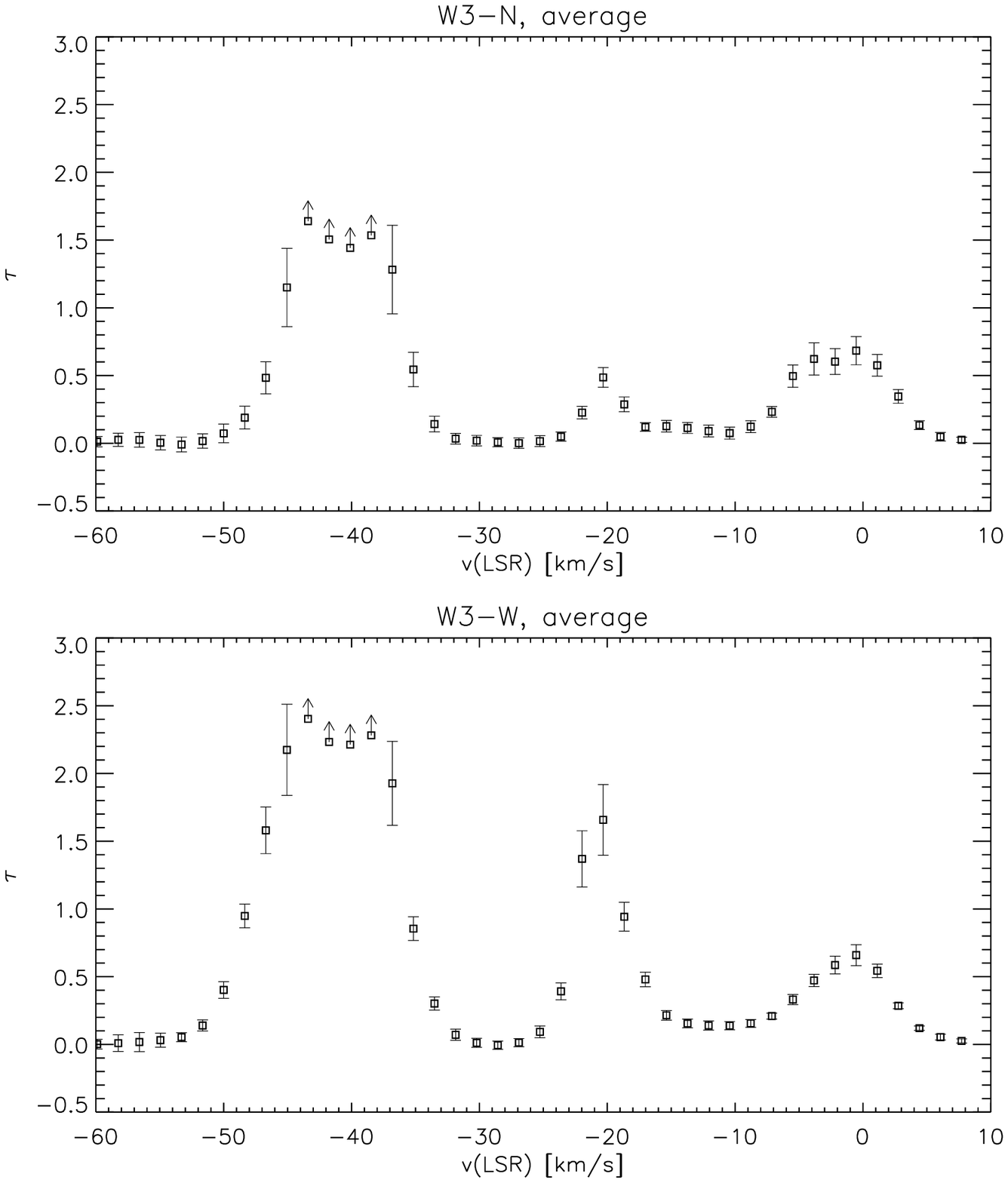}

\figcaption[Normandeau.fig4.ps]{The optical depth profiles towards
	W3-N and W3-W.
	These were calculated using the on-source profiles shown in
	Fig.~\ref{fig:W3_abs_em}, i.e.\ the average spectra towards
	W3-N and W3-W, and the off-source spectra shown in
	Fig.~\ref{fig:ave_em}. The points for which lower limits are
	indicated correspond to velocities for which $T_{\em off} -
	T_{\em on}$ is within $2(\sigma_{\em off} +
	\sigma_{\em on})$ of $T_c$.\label{fig:tau}}

\clearpage

\plotone{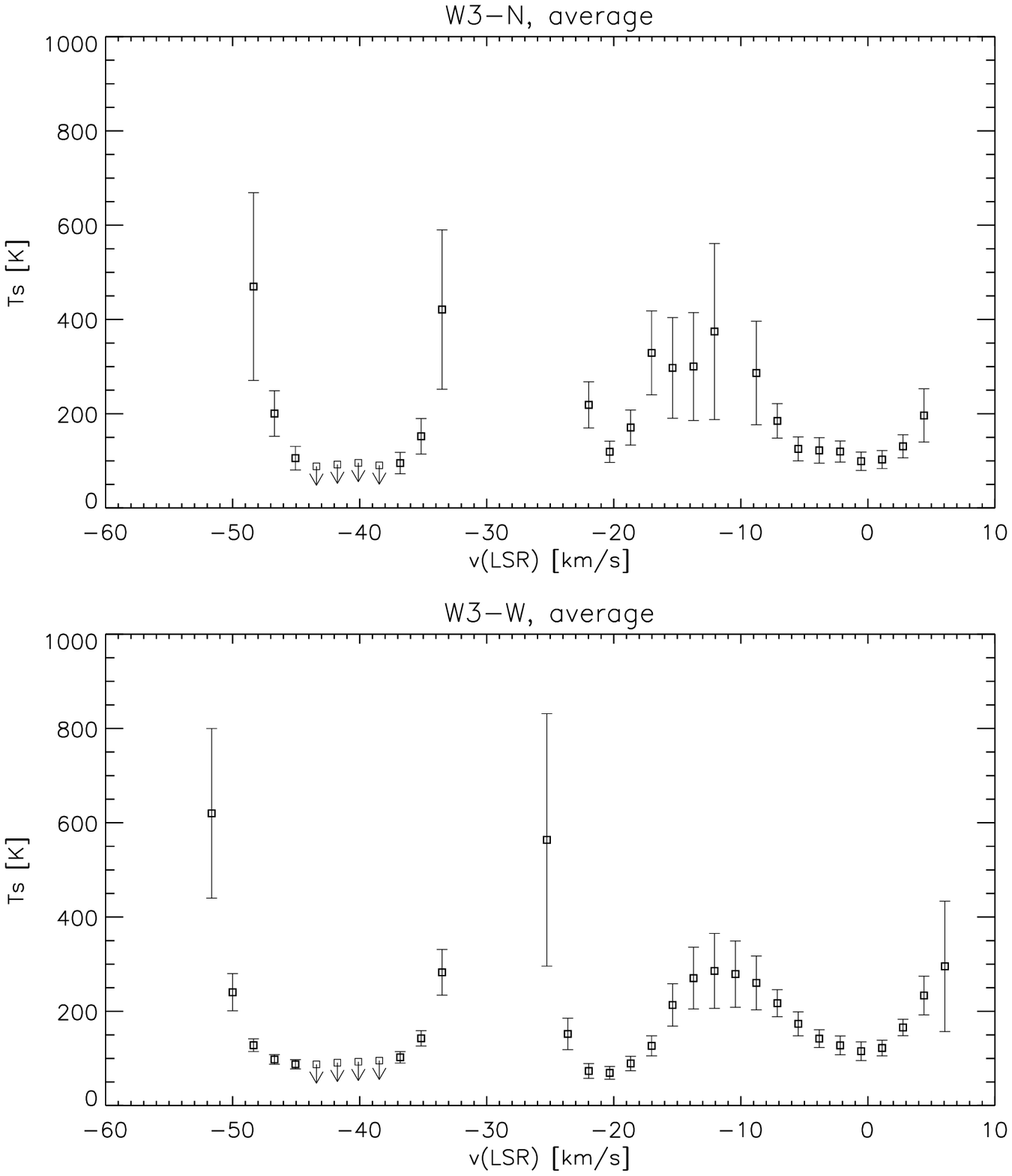}

\figcaption[Normandeau.fig5.ps]{The spin temperature profiles towards
	W3-N and W3-W.
	These were calculated using the optical depth values plotted
	in Fig.~\ref{fig:tau}, and the off-source spectra shown in
	Fig.~\ref{fig:ave_em}. The points for which upper limits are
	indicated correspond to velocities for which only lower limits
	were calculated for $\tau$. Data points are only plotted for
	those channels where the calculated spin temperature was more
	than twice its uncertainty.\label{fig:Tspin}} 

\clearpage

\plotone{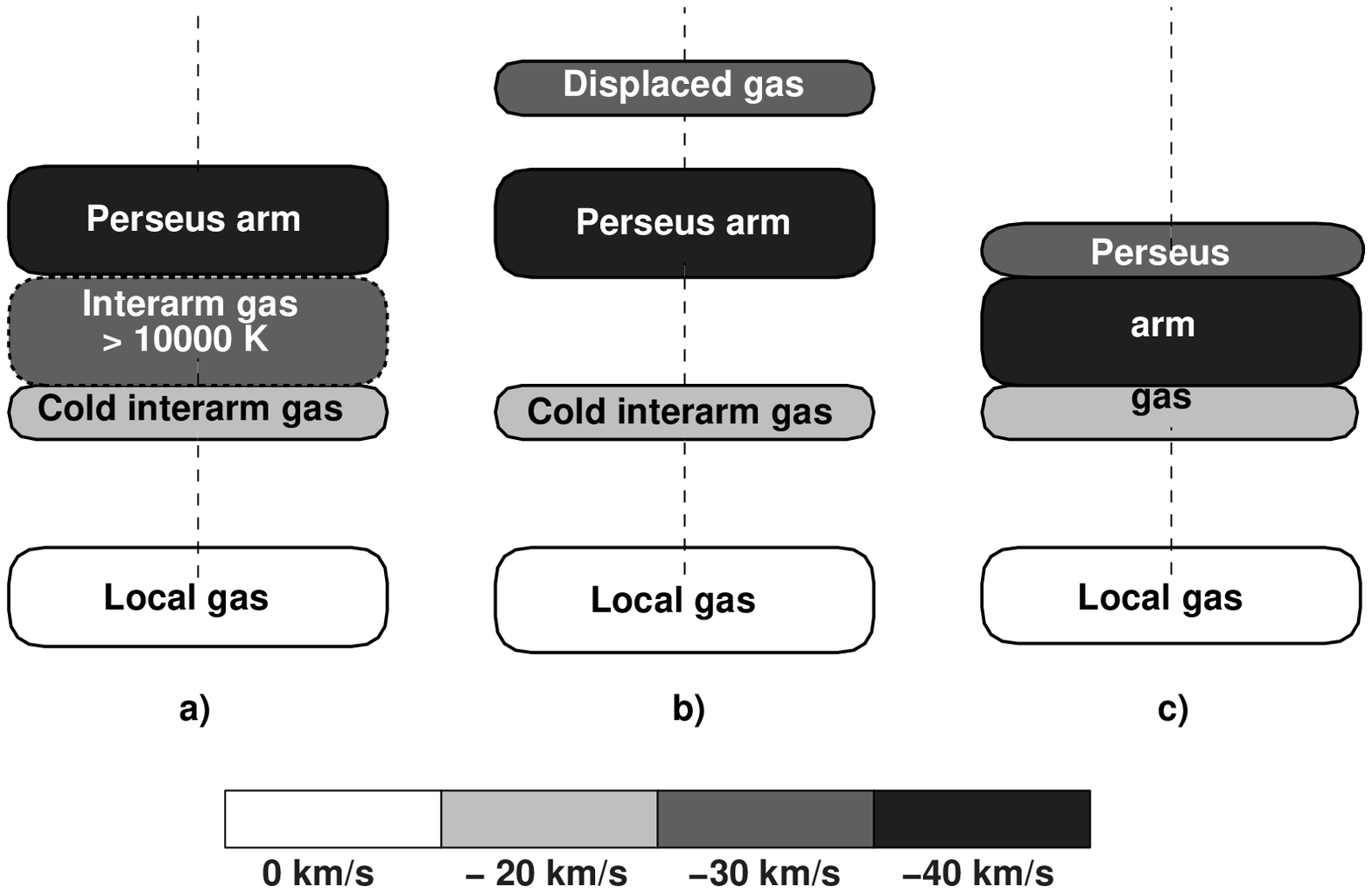}

\vspace{2cm}

\figcaption[Normandeau.fig6.ps]{Schematic representation of the three 
	explanations considered for the lack of absorption at --30 \kms.
	The diagrams are not to scale.
	a) There is no absorption by the interarm gas around --30 \kms\
	even though it is in front of W3. This implies spin temperatures
	in excess of $10^4$ K which is unlikely for warm, neutral gas. 
	b) The gas at --30 \kms\ is behind W3 but is not following the 
	rotation curve. 
	c) The gas at --40 \kms\ is shocked gas in the Perseus arm whereas
	the gas at --30 \kms\ is unshocked gas in the Perseus arm. This is
	the picture suggested by the TASS model (Roberts
	1972).\label{fig:diag}} 

\end{document}